\documentclass[12pt,preprint]{aastex}

\usepackage{amsmath}

\newcommand{\amm}{NH$_{3}$}
\newcommand{\mh}{H$_{2}$}

\newcommand{\cden}{cm$^{-2}$}
\newcommand{\vden}{cm$^{-3}$}

\newcommand{\msun}{M$_{\sun}$}
\newcommand{\kms}{km~s$^{-1}$}

\slugcomment{}

\shorttitle{Ammonia in Arp\,220}
\shortauthors{Ott et al.}

\begin{document}

\title{Ammonia as a Temperature Tracer in the Ultraluminous
  Galaxy Merger Arp\,220}

\author{J{\"u}rgen Ott}
\affil{National Radio Astronomy Observatory, P.O. Box O, 1003
  Lopezville Road, Socorro, NM 87801, USA}
\email{jott@nrao.edu}

\author{Christian Henkel}
\affil{Max-Planck-Institut f{\"u}r Radioastronomie, Auf dem H{\"u}gel
  69, 53121 Bonn, Germany} 
\affil{Astronomy Department, Faculty of Science,
King Abdulaziz University, P.O. Box 80203,
Jeddah, Saudi Arabia}
\email{chenkel@mpifr-bonn.mpg.de}

\author{James A. Braatz}
\affil{National Radio Astronomy Observatory, 520 Edgemont Road,
  Charlottesville, VA 22903, USA}
\email{jbraatz@nrao.edu}

\and

\author{Axel Wei{\ss}}
\affil{Max-Planck-Institut f{\"u}r Radioastronomie, Auf dem H{\"u}gel 69, 53121 Bonn, Germany} 
\email{aweiss@mpifr-bonn.mpg.de}

\begin{abstract}

  We present Australia Telescope Compact Array (ATCA) and Robert
  C. Byrd Green Bank Telescope (GBT) observations of ammonia (\amm)
  and the 1.2\,cm radio continuum toward the ultraluminous infrared
  galaxy (ULIRG) merger Arp\,220.  We detect the \amm (1,1), (2,2),
  (3,3), (4,4), (5,5), and (6,6) inversion lines in absorption against
  the unresolved, $(62\pm9)$\,mJy continuum source at 1.2\,cm.  The
  peak apparent optical depths of the ammonia lines range from $\sim$
  0.05 to 0.18.  The absorption lines are well described by
  single-component Gaussians with central velocities in between the
  velocities of the eastern and western cores of Arp\,220. Therefore,
  the ammonia likely traces gas that encompasses both cores. The
  absorption depth of the \amm (1,1) line is significantly shallower
  than expected based on the depths of the other transitions. The
  shallow (1,1) profile may be caused by contamination from emission
  by a hypothetical, cold ($\lesssim 20$\,K) gas layer with an
  estimated column density of $\lesssim 2\times10^{14}$\,\cden. This
  layer would have to be located behind or away from the radio
  continuum sources to produce the contaminating emission. The widths
  of the ammonia absorption lines are $\sim 120-430$\,\kms, in
  agreement with those of other molecular tracers. We cannot confirm
  the extremely large linewidths of up to $\sim 1800$\,\kms\
  previously reported for this galaxy.  Using all of the ATCA
  detections except for the shallow (1,1) line, we determine a
  rotational temperature of $(124\pm19)$\,K, corresponding to a
  kinetic temperature of $T_{\rm kin}=(186\pm55)$\,K. Ammonia column
  densities depend on the excitation temperature. For excitation
  temperatures of 10\,K and 50\,K, we estimate $N({\rm
  NH_3})=(1.7\pm0.1)\times10^{16}$\,\cden\ and
  $(8.4\pm0.5)\times10^{16}$\,\cden, respectively. The relation scales
  linearly for possible higher excitation temperatures. Our
  observations are consistent with an ortho--to--para-ammonia ratio of
  unity, implying that the ammonia formation temperature exceeds $\sim
  30$\,K. In the context of a model with a molecular ring that
  connects the two nuclei in Arp\,220, we estimate the \mh\ gas
  density to be $\sim f_{\rm V}^{-0.5} \times (1-4)\times 10^{3}$,
  where $f_{\rm V}$ is the volume filling factor of the molecular gas.
  In addition to ammonia, our ATCA data show an absorption feature
  adjacent in frequency to the \amm(3,3) line.  The line does not
  appear in the GBT spectrum.  If we interpret the line to be from the
  OH $^{2}\Pi_{3/2}$ $J=9/2$ $F=4-4$ transition, it would have a
  linewidth, systemic velocity, and apparent optical depth similar to
  what we detect in the ammonia lines.  Comparing the new line to the
  previously detected 6\,GHz OH $^{2}\Pi_{3/2}$ $J=5/2$ $F=2-2$
  transition, we determine a rotational OH temperature of $\sim
  245$\,K, about 2 times the rotational temperature of ammonia.  If
  this association with OH is correct, it marks the first detection of
  the highly excited ($\sim 511$\,K above ground state)
  $^{2}\Pi_{3/2}$ $J=9/2$ $F=4-4$ OH line in an extragalactic object.

\end{abstract}

\object{Arp 220}


\section{Introduction}
\label{sec:intro}

Ultraluminous infrared galaxies (ULIRGs), with far--infrared
luminosities exceeding $L_{\rm FIR}\ga 10^{12}$\,L$_{\odot}$, exhibit
some of the highest star formation (SF) rates encountered in the
Universe \citep[for a review, see][]{san96}. They are mergers
containing at least two massive, gas-rich galaxies whose dust is
heated during the merging process. Cloud-cloud collisions and the
subsequent inflow of gas feed nuclear activity \citep[][]{nag03} and
trigger an amazingly high rate of star formation in the central few
100\,pc, typically reaching tens to hundreds of M$_{\odot}$\,yr$^{-1}$
\citep[e.g., ][]{flo04}. ULIRGs are particularly important galaxies as
they are the most readily detectable population for which the SF per
co--moving volume can be traced out to high redshifts
\citep[e.g.,][]{mad98,col01,dar02}. Furthermore, the number of
luminous IR galaxies per co-moving volume was much greater at high
redshift, so these galaxies dominated the production of stars early in
the Universe \citep[e.g.][]{bri07,hop10}. There is increasing evidence
that ULIRGs may enter an optically bright quasar phase subsequently in
the merging process \citep[e.g.][]{hut88,can01,das07,vei09}.

The prototypical ULIRG is Arp\,220 (IC\,4553; $L_{\rm
FIR}\sim1.5\times10^{12}\,L_{\odot}$), at a distance of only $D \sim$
75\,Mpc. Its proximity allows us to study physical processes in more
detail than we can for its more distant cousins. Arp\,220 has two
starburst nuclei \citep[e.g.,][]{nor88,dow98} separated by 0\farcs95
(350\,pc) at a position angle of 92$^{\circ}$. Many molecules and
associated tracers have been detected in Arp\,220, including
CO,\ion{C}{1}, \ion{C}{2}, CS, HCN, HNC, HCO$^{+}$, H$_{2}$CO,
H$_{2}$O,H$_{2}$S, HC$_{3}$N, HNCO, and thermal
OH\citep[e.g.,][]{sol90,sol92,hue95a,sco97,sol97,ger98,pap98,yao03,ara04,dow07,sal08,aal09,gre09,mar11}
as well as OH megamaser emission \citep[e.g., ][]{baa82,rov03}.  The
molecular gas traced by (sub-)arcsecond CO imaging reveals small,
dense disks around each of the two nuclei.  These small disks are
embedded in a ring or disk that encompasses both cores, and that ring
is itself embedded in an even larger molecular disk, $\sim 2.5$\,kpc
in size and with a density of $\sim10^{2}$\,\vden\
\citep{sco97,dow98,sak99,mun01}. The western nucleus has a higher
molecular mass than its eastern counterpart, and may contain a
supermassive black hole \citep{dow07}. Both nuclei show signatures of
outflows \citep{sak09}. Direct measurements of the physical properties
of the gas, however, are difficult. For linear molecules, radiative
transfer models, like the Large Velocity Gradient (LVG) analysis, have
solutions that are degenerate between high temperature/low density and
low temperature/high density.  Analysis of the excitation conditions
show that the two cores have temperatures in the range $\sim
45-120$\,K and densities of order 10$^{6}$\,\vden, with the western
nucleus being somewhat warmer and denser
\citep[e.g.][]{aal09,gre09,mat09}.  Farther out, while the gas is
known to be warm ($T\gtrsim40$\,K), no accurate temperature or density
measurements exist.

\citet{sol92} and \citet{gao04a,gao04b} observed CO and HCN in
a large sample of ULIRGs. They conclude that HCN is a much better
tracer of SF than CO, and in fact they find that the HCN line
luminosity scales linearly with the SF rate in galaxies. HCN
exhibits a large electric dipole moment and the molecule is only
excited when the gas is sufficiently dense ($n$(H$_2$) $\gtrsim$
10$^4$\,cm$^{-3}$). Dense molecular gas is therefore the best tracer
for the material that is readily converted into stars \citep[but
see][for caveats]{gar06}.

Similar to HCN, ammonia (NH$_3$) also traces dense gas, though it is
excited at densities down to $\sim 10^{3-4}$\,\vden.  Its abundance is
closely linked to that of HCN by the cyanide chemistry network
\citep[e.g., ][]{sch92}. In contrast to linear molecules like CO and
HCN, the ammonia molecule is a prolate symmetric top with tetrahedral
structure. Most of its excitation states, characterized by the quantum
numbers ($J$,$K$), show `inversion doublets' caused by the nitrogen
atom tunneling through the plane defined by the three hydrogen
atoms. The lowest inversion doublet of each $K$-ladder (with the
quantum numbers $J$=$K$) is metastable. The relative populations of
the metastable doublets are mainly determined by collisions, follow a
Boltzmann distribution, and thus provide one of the most suitable
thermometers to estimate kinetic temperatures of extragalactic
molecular gas \citep[e.g.,][]{wal83,ung86,dan88,hen00}. Since the
energy separation between the individual inversion doublets is only a
weak function of $J$ and $K$, a vast range of excitation properties
can be covered by observations of lines closely spaced in frequency.
The most relevant NH$_3$ lines fall in the 1.2\,cm band
($\nu\sim25$\,GHz).  Relative calibration of these lines tends to be
very good since they can be observed nearly simultaneously, under
similar atmospheric conditions, and with the same telescope, receiver,
and backend.

\amm\ has been detected in Arp\,220 by \citet{tak05} with the 45\,m
Nobeyama single dish telescope. Their main result was that the ammonia
has linewidths of $\sim 1800$\,\kms, which they suggest could indicate
that the gas is located in an accretion disk around an active
nucleus. Alternatively, the wide lines may indicate strong molecular
inflows or outflows. Single dish observations, however, can be
affected by instabilities that lead to poor spectral baselines and
uncertain line profiles. This is particularly the case for broad
lines. Here, we present ATCA interferometric data with higher signal
to noise and better spectral baselines.  We also present a sensitive
GBT spectrum. In Sect.\,\ref{sec:obs} we describe our observational
setups. We present our results in Sect.\,\ref{sec:results}, along with
a critical comparison of our data with those of \citet{tak05}. In
Sect.\,\ref{sec:discuss} we derive and interpret the apparent optical
depths, gas temperatures, gas densities and kinematics of Arp\,220, as
well as its SF properties.  Sect.\,\ref{sec:summary}
summarizes our paper.

\section{Observations and Data Reduction}
\label{sec:obs}

\subsection{ATCA Observations}
\label{sec:obsATCA}

We observed the spectral lines of ammonia and the 1.2\,cm (K-band)
continuum toward Arp\,220 with the ATCA\footnote{The Australia
Telescope Compact Array is part of the Australia Telescope which is
funded by the Commonwealth of Australia for operation as a National
Facility managed by CSIRO.}. The data were taken in several stages:
(1) A broadband spectrum was obtained covering the 23.2-23.6\,GHz
frequency range. At the time of the observations (2005 July), the ATCA
could not cover this full frequency range in a single
observation. Therefore, we configured the observations with 128\,MHz
bands of 64 channels each, and stepped through the desired frequency
range in 60\,MHz increments (indicated by the bars at the bottom of
Fig.\,\ref{fig:atcabroad}), with two bands observed at a time. The
observations were performed over 4 days with $\sim 4.5$\,h on-source
per day. The spectral resolution is equivalent to 2.2 channels ($\sim
57$\,\kms).  Calibration was performed individually for each frequency
band and observing day. At the time, the ATCA was in its most compact
H\,75 antenna configuration, where baselines reach from 31\,m to
89\,m. (2) The (6,6) line was observed in 2005 March with the H\,214
antenna configuration, which has 82\,m-247\,m baselines.  The
integration times on source were $\sim 2.5$\,h (2005 March 17), and
$\sim 4.5$\,h (2005 March 20).  (3) On 2005 July 30 we observed the
(4,4) line on source for 4.5\,h in the H\,75 configuration. (4) A day
later we measured the (5,5) line with the same integration time in the
same array configuration. The correlator settings for stages (2)-(4)
were the same as in stage (1), except that, for the (4,4) and (5,5)
transitions, only two 128\,MHz wide frequency settings were used in
parallel and separated by 72\,MHz. The (6,6) line was observed in a
single frequency window. A summary of the observational setups is
provided in Table\,\ref{tab:obs}.

The data were processed using the {\sc MIRIAD} software package.  We
observed PKS\,1253--055 for bandpass calibration, PKS\,1934--638 for
flux calibration, and PKS\,1538+149 (in 20\,min intervals) for
phase/gain calibration. For each spectrum, we subtracted the continuum
as determined by a linear fit to the line-free channels.  In the
process we also created continuum data sets. All data were Fourier
transformed using `natural' weighting.  To obtain the spectra, the
data were not deconvolved due to the lack of spatial structure on the
scales of the dirty beam (source size $\ll 10\arcsec$ versus $\sim
10\arcsec-40\arcsec$ beam size, the latter depending on the array
configuration, see Table\,\ref{tab:obs}). Deconvolution with {\sc
Clean} is only needed to reconstruct source morphologies. This,
however, comes for the prize of introducing non-linearities. For our
unresolved source, we decided not to {\sc Clean} the data. Given the
relatively strong continuum flux, this implies that the noise within
the planes of the dirty images is dominated by sidelobes and we
measure $5-10$\,mJy\,beam$^{-1}$ depending on the observations. For an
estimate of the more representative statistical noise and thus for the
apparent optical depth uncertainty, however, we did apply {\sc Clean}
deconvolution to remove the sidelobes. This method yields rms values of $\sim
0.4-1.2\,$mJy\,beam$^{-1}$ per channel for the different observations.
We estimate the absolute calibration uncertainties to be of order
15\%.

\subsection{GBT Observations}
\label{sec:obsGBT}
We also observed the 1.2\,cm ammonia spectrum toward Arp\,220 on 2006
January 21 and 2006 February 02 with the Green Bank Telescope (GBT) of
the National Radio Astronomy Observatory\footnote{The National Radio
Astronomy Observatory is a facility of the National Science Foundation
operated under cooperative agreement by Associated Universities, Inc.}
(see Table\,\ref{tab:obs}). We observed both circular polarizations
simultaneously using a total-power nodding mode in which the target
position was placed alternately every 2.5\,min in one of the two beams of the
$22-26$\,GHz K-band receiver. The beams are spaced by
3\arcmin\ in the cross-elevation direction. Total time on source
was 40 minutes. We configured the spectrometer with an 800\,MHz
spectral window centered on the \amm(3,3) line for Arp\,220.  The data
were calibrated and averaged in GBTIDL.  During calibration, we used
an estimate of the atmospheric opacity obtained from weather
conditions.  We boxcar averaged the spectrum to achieve 1.56\,MHz
channel spacing, and subtracted a fifth-order polynomial fit to the
line-free channels over the full 800\,MHz, a conservative approach for
broadband GBT data.  Finally we combined the spectra from the two days
and the rms noise of the final data is $0.8$\,mJy per $1.56$\,MHz
channel.

\section{Results}
\label{sec:results}

\subsection{1.2\,cm Continuum Emission}
\label{sec:cont}
With the ATCA, we detect an unresolved continuum source
toward Arp\,220 at RA\,(J2000) = $15^{h}\,34^{m}\,57.2^{s}$,
DEC\,(J2000) = $23\degr\,30\arcmin\,11\arcsec$.  The source has a flux density
at 23.4\,GHz of 208\,mK or $62$\,mJy (error of order 15\%) measured
within the $31.1\arcsec\times21.1\arcsec$ beam.
From March to August 2005, the measured continuum flux 
varied by less than 5\%. 

\subsection{Ammonia Inversion Lines}
 
\begin{figure}
\epsscale{1}
\plotone{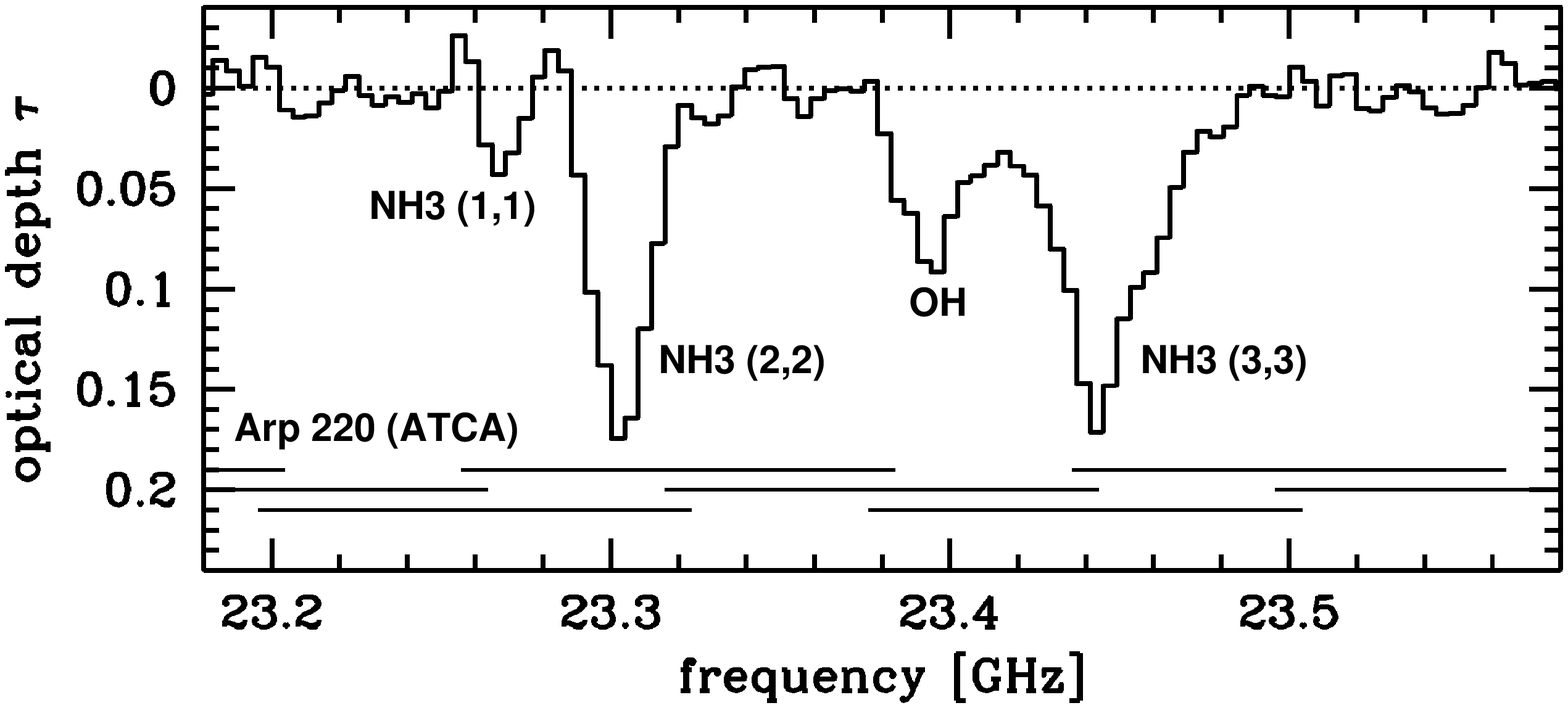}
\caption{ATCA broadband spectrum in the 23.2 to 23.6\,GHz range. Units
  are in apparent optical depth $\tau$. The bars at the bottom indicate the
  bandwidths and tunings of the different spectral settings. The rms
  noise of the observations is $\tau_{\rm rms}\approx 0.013$. \label{fig:atcabroad}}
\end{figure}

\begin{figure}
\epsscale{1}
\plotone{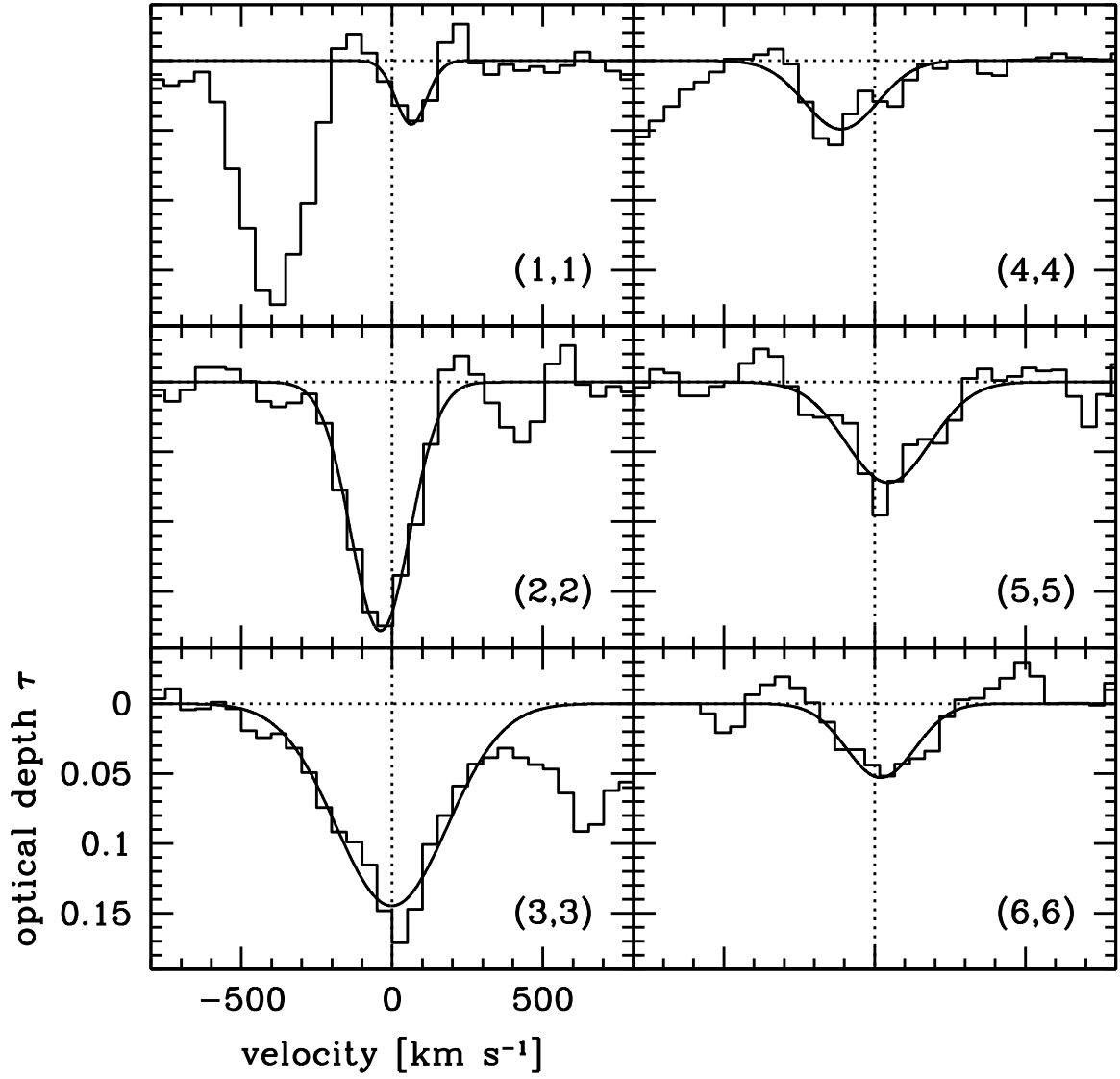}
\caption{
Continuum subtracted ammonia absorption ATCA spectra of Arp\,220 in units
of apparent optical depth $\tau$. Gaussian fits to the lines are overlaid. 
All panels are on the same scale. \label{fig:spec}}
\end{figure}

\begin{figure}
\epsscale{1}
\plotone{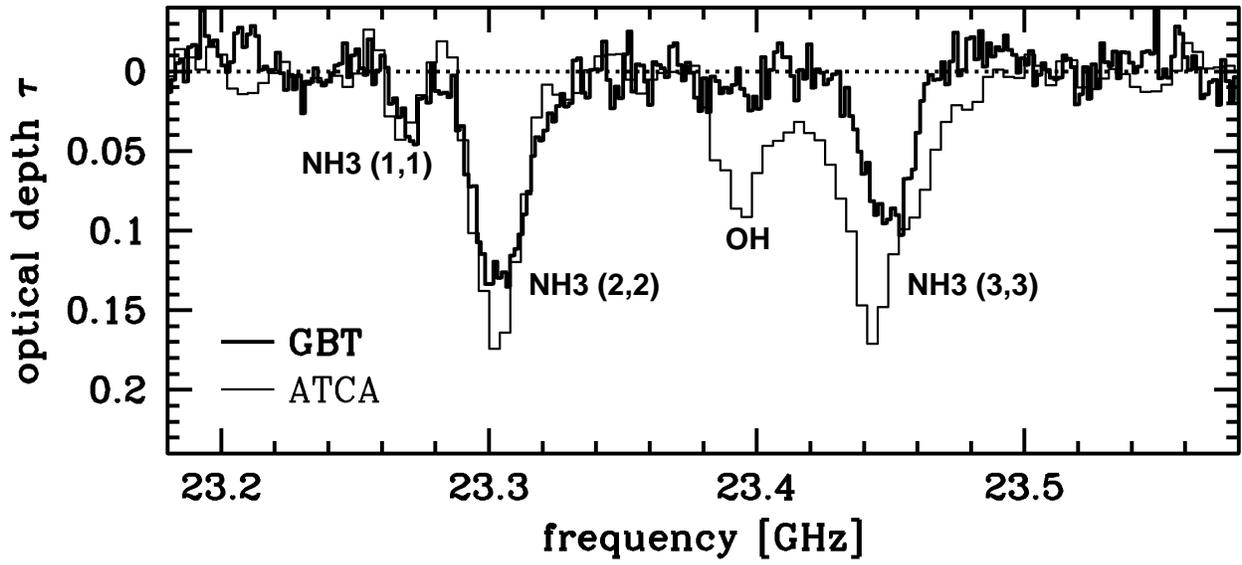}
\caption{
GBT broadband spectrum (thick lines) overlaid over the ATCA spectrum.\label{fig:gbtspec}} 
\end{figure}

In Figs.\,\ref{fig:atcabroad} and \ref{fig:spec} we show ATCA spectra
of Arp\,220 in units of apparent optical depth (see
Sect.\,\ref{sec:tau}), and Fig.\,\ref{fig:gbtspec} shows a GBT
spectrum overlaid on the ATCA data. All six observed ammonia inversion
lines, NH$_3$ ($J$,$K$) = (1,1), (2,2), (3,3), (4,4), (5,5), and
(6,6), are detected in absorption against the radio continuum of
Arp\,220.  We do not detect clear emission features, but the broadband
ATCA spectrum displays slightly positive values adjacent to the (1,1)
absorption line (see Sects.\,\ref{sec:tau} and \ref{sec:T} for a
discussion). We fitted single-component Gaussian profiles to all
observed absorption lines. The resulting apparent optical depths
$\tau$ (see Sec.\,\ref{sec:tau} for their derivation), the optical,
barycentric peak velocities $v_{\rm opt,bary}^{\rm peak}$, the
linewidths $\Delta v_{1/2}$, and the apparent optical depths
integrated over the line profiles, $\int \tau\, $d$v$, are given in
Table\,\ref{tab:line} and the fits are shown in Fig\,\ref{fig:spec}.
The fitted line centroids are all within $\sim 60$\,\kms\ of the
systemic velocity of $5434$\,\kms\ ($z=0.018126$). This is remarkable
given the $\sim 57$\,\kms\ velocity resolution of the ATCA
observations (see Sect.\,\ref{sec:obs}). The linewidths range between
$\sim 230$\,\kms\ and $\sim 330$\,\kms\ for most lines. The (1,1) and
(3,3) lines, however, have widths of $121\pm38$\,\kms\ and
$435\pm19$\,\kms, more extreme than the other transitions.  Note,
though, that the (3,3) line slightly overlaps with another absorption
feature, described in Sect.\,\ref{sec:resultsOH}. Our fitted
linewidths are in agreement with those found for other molecular
tracers \citep[e.g.][]{sal08} but contradict the results presented in
\citet{tak05} for the \amm(1,1) and (3,3) lines. We cannot confirm the
extremely wide lines of up to $\sim 1800$\,\kms\ observed by
\citet{tak05} with the Nobeyama 45\,m telescope.  None of the lines
measured with the ATCA or GBT have widths exceeding $\sim 450$\,\kms\
(see Table\,\ref{tab:line} and
Figs.\,\ref{fig:atcabroad}--\ref{fig:gbtspec}).

There are some notable differences in widths and strengths of the
lines measured with the GBT and the ATCA. The reason may be the
weakness of the lines, the complexity of the spectra, and potential
confusion with the continuum emission from Arp220. The \amm(1,1) and
(2,2) lines are separated by less than 30\,MHz so, for the GBT data,
it is difficult to precisely define a spectral baseline near these
features. The quality of spectral baselines from the interferometric
ATCA data, on the other hand, is very good.  We would like to add,
however, that regridding the 128\,MHz blocks of our ATCA observations
into a single spectrum conveys some uncertainty. For the relatively
isolated \amm(4,4) line, our results agree well with those of
\citet{tak05}.  The centroids of the Gaussian line fits between the
GBT and ATCA also differ by a few tens of \kms\
(Table\,\ref{tab:line}). This may be caused by the limited
signal-to-noise ratios of the spectra, despite the relatively small
errors of the Gaussian fits. Gaussian fitting may have small
systematic errors caused by the uncertain baseline subtraction and
blended lines.  The linewidths and peak apparent optical depths of the
\amm(1,1) line agree between the ATCA and GBT observations. The other
lines, however, disagree to some extent in linewidth and peak apparent
optical depth. The (3,3) line is about twice as broad in the ATCA data
as in the GBT data and the peak apparent optical depths in the GBT
data for the \amm(2,2) and (3,3) lines are about 1/3 less than those
in the ATCA spectrum.

\subsection{Potential Detection of OH $^{2}\Pi_{3/2}$ $J=9/2$}
\label{sec:resultsOH}

In addition to the (1,1) through (6,6) ammonia absorption lines, the
broadband ATCA spectrum exhibits an additional absorption feature at
$\sim 23.39$\,GHz (Fig.\,\ref{fig:atcabroad}).  As with the ammonia
lines, this additional spectral line appears in both spectra that were
taken at that frequency (the horizontal bars in
Fig.\,\ref{fig:atcabroad} show the tunings of the individual
observations).  Note, however, that the two underlying individual
spectra were observed simultaneously which implies that they are
subject to a common systematic uncertainty and are not entirely
independent measurements. At the redshift of Arp 220, the most obvious
line candidates would be the $F$=4-4 and $F$=5-5 doublet of the OH
$^{2}\Pi_{3/2}$ $J=9/2$ transition at rest frequencies of
23817.6153\,MHz and 23826.6211\,MHz, respectively\footnote{taken from
the LOVAS catalog: {\tt
http://physics.nist.gov/cgi-bin/micro/table5/start.pl }}. This doublet
is at an energy $\sim 511$\,K above ground state and indicates highly
excited gas.  Adopting the rest frequency of the F=4-4 transition and
fitting a Gaussian to this line, we get a line center of
$-13\pm10$\,\kms\ (relative to the systemic velocity of $5434$\,\kms)
and a linewidth of $292\pm29$\,\kms, very similar to the ammonia
lines.  These fitting results are also shown in Table\,\ref{tab:line}.
Fitting the F=5-5 line alone, we find the centroid of the Gaussian
would be offset by $\sim +100$\,\kms\ from systemic, likely too large
for a feature that belongs to Arp 220.  Since the doublet lines are
spaced by only $\sim 113$\,\kms, however, they may be blended to
produce the observed line width.  A fit from blended lines would have
a mean velocity offset of $\sim +44$\,\kms\ -- well within the range
of the line velocities observed. In either case, if the measured
absorption feature indeed corresponds to one or both OH
$^{2}\Pi_{3/2}$ $J=9/2$ lines, it would be the first detection of this
transition in an extragalactic object. To date, this doublet has only
been observed toward Galactic ultracompact \ion{H}{2} regions such as
W3(OH) \citep[e.g.][]{win78,wil91,bau95}.

Lower energy OH doublets have been observed previously in Arp\,220.
These include $^{2}\Pi_{1/2}$ $J=1/2$ at 4.7\,GHz, $^{2}\Pi_{3/2}$
$J=5/2$ at 6.0\,GHz, and $^{2}\Pi_{1/2}$ $J=3/2$ at 7.8\,GHz
\citep{hen86,hen87,sal08}. These detections show offset velocities in
the range $-20$\,\kms\ to $+5$\,\kms, linewidths of $\sim
250-480$\,\kms, and apparent optical depths of $\tau\sim0.05-0.11$.
The parameters of these lines are consistent with those of our
possible ATCA OH line detection.  Using our possible line and the
\citealt{sal08} $^{2}\Pi_{3/2}$ $J=5/2$ F=2-2 detection, we estimate a
rotational temperature of $\sim 245$\,K, assuming Local Thermal
Equilibrium.  This estimate is about a factor of two higher than the
rotational temperature we determine for ammonia (Sect.\,\ref{sec:T}).
The line could be verified with wideband interferometric observations,
for example with the new CABB backend of the ATCA or WIDAR on the
EVLA. Indirect evidence could come from {\it HERSCHEL} data, via
observations of rotational lines between different OH doublets
\citep[e.g.][]{wam10}.

\section{Discussion}
\label{sec:discuss}

\subsection{Apparent Optical Depths}
\label{sec:tau}

For a molecular cloud that partly covers a continuum source, 
in the Rayleigh--Jeans limit, we can express the measured 
brightness temperature of a spectral line ($T_{\rm L}$) as:

\begin{equation}
T_{\rm L} = (f_{\rm cl}\, T_{\rm ex} - f_{0}\, f_{\rm C}\, T_{\rm C} -
f_{\rm cl}\, T_{\rm CMB}) \times\ (1-e^{-\tau})
\label{eq:taua}
\end{equation}

\noindent \citep{roh04} where $T_{\rm ex}$ is the excitation
temperature of the line, $T_{\rm C}$ is the brightness temperature of
a background continuum source, $T_{\rm CMB}$ is the brightness
temperature of the cosmic microwave background, and $\tau$ is the
optical depth of the line. The temperatures are scaled by the
beam filling factors of the cloud ($f_{\rm cl}$) and the continuum
source ($f_{\rm C}$).  The source covering factor $f_{0}$ describes
the fraction of the continuum covered by the cloud along the line of
sight. We use this equation to determine the apparent optical depth
$\tau$. The source covering factor $f_{0}$ can be assumed to be close
to unity since the individual radio continuum emitting regions are
relatively small, at most a few tens of pc in size
\citep{rov03,lon06,par07}, and supposedly contain mostly supernova
remnants embedded in the molecular gas (see Sect.\,\ref{sec:kin}).  In
addition, the mean gas column density in the model of \citet{dow98} is
$N_{H_{2}}\sim 2.5\times10^{24}$\,\cden\ and the mean gas volume
density is $\sim 500$\,\vden\ (using the radius of the outer molecular
disk of 1.2\,kpc, which corresponds for a cylindrical, edge-on
geometry to a mean width and line of sight of 1.6\,kpc, a height of
the cylinder of 80\,pc, and a gas mass of $5\times10^{9}$\,\msun; but
see Sect.\,\ref{sec:dens}). Those large densities indicate that
molecular gas is abundant and, closer to the starburst cores, ammonia
is expected to be present over large regions which justifies our
assumption of a source covering factor near unity. Under these
circumstances we can also assume that both the cloud and the continuum
filling factors are the same for absorption lines, i.e., $f_{\rm
cl}=f_{\rm C}$. Ammonia not along the line of sight toward the radio
continuum would appear in emission, which is not directly observed
(but note the weak, tentative emission features adjacent to the
\amm(1,1) line [Fig.\,\ref{fig:atcabroad}] -- the peculiarity of the
(1,1) transition is discussed in Sect.\,\ref{sec:T}). Under these
conditions, Eq.\,\ref{eq:taua} simplifies to

\begin{equation}
T_{\rm L} = f_{\rm C}(T_{\rm ex} - T_{\rm C} - T_{\rm CMB}) \times\ (1-e^{-\tau}).
\label{eq:taub}
\end{equation}

VLBI continuum measurements at 18\,cm show that the radio continuum of
Arp\,220 has an extent of $\lesssim 0.6\square\arcsec$
\citep{smi98,rov03}, which corresponds to a beam filling factor of
$f_{\rm C}\lesssim 10^{-3}$ when compared to the beam of our ATCA
broadband observations ($\sim31\farcs1\times21\farcs1$). For this
value, the measured Arp\,220 1.2\,cm continuum brightness temperature
of $T_{\rm C}'= f_{\rm C}\, T_{\rm C}= 208$\,mK then implies a true
brightness temperature $T_{\rm C}\gtrsim208$\,K.  The best spatial
resolution was obtained for the \amm(6,6) line
(Sect.\,\ref{sec:obsATCA}) and even for the $\sim 8$ times (in area)
smaller beam of that observation we do not detect any extent of the
continuum source. For this smaller beam of the \amm(6,6) observation,
the same approximation then boosts the true continuum temperature
limit by the same factor of 8. On the other hand, VLBI does filter out
flux on larger scales which somewhat counters the low beam filling
argument. Nevertheless, the true continuum brightness temperature is
very likely much larger than the CMB temperature of $T_{\rm
  CMB}=2.73$\,K. We can therefore neglect $T_{\rm CMB}$ in all further
calculations. We can also disregard $T_{\rm ex}$. LVG models of the
ammonia molecule \citep[e.g.,][]{wal83} show that, under normal
circumstances, the excitation of inversion lines hardly exceeds
$T_{\rm ex}\sim 50$\,K (see also Sect\,\ref{sec:dens}). The continuum
brightness temperature therefore dominates the radiative
transfer. With those assumptions, Eq.\,\ref{eq:taub} can be expressed
as

\begin{equation}
\tau = -\ln \left(1-\frac{|T_{\rm L}|}{T_{\rm C}'}\right), 
\label{eq:tau}
\end{equation}
with the approximated $\tau$ denoting the 'apparent optical depth'.

In Figs.\,\ref{fig:atcabroad}, \ref{fig:spec}, and \ref{fig:gbtspec}
the absorption profiles of the ATCA and GBT observations are displayed
in units of apparent optical depth (using for both the ATCA continuum flux described
in Sec.\,\ref{sec:cont}). The Gaussian fits to the spectra
(Fig.\,\ref{fig:spec}) result in peak apparent optical depths of $\sim
0.05-0.18$ (Table\,\ref{tab:line}) which indicate optically thin
lines. The peak apparent optical depths are similar to what \citet{tak05} finds
for the \amm(1,1), (3,3), and (4,4) lines. Our fit to the (2,2) line,
however, is about 3 times larger than theirs. We should also note that
the integrated apparent optical depths are different due to the larger
linewidths reported by \citet{tak05}.

\subsection{Rotational and Kinetic Temperature}
\label{sec:T}

Using the apparent optical depths, it is possible to derive the ratios of the
ammonia column densities $N(J,K)$ and excitation temperatures $T_{\rm
ex}$ for the two states of a given inversion doublet with the equation

\begin{equation}
 \frac{N(J,K)}{T_{\rm ex}} = 1.61 \times 10^{14} \times\ \frac{J(J+1)}{K^2\,\nu} \times\ \tau \times\ \Delta v_{1/2}
\label{eq:NT}
\end{equation}
(\citealt{hue95b}; $\nu$ in GHz; $\Delta v_{1/2}$: full width half
maximum [FWHM] linewidth in km\,s$^{-1}$). Normalizing these with
respect to their statistical weights and assuming that the excitation
temperatures $T_{\rm ex}$ are the same for the six observed
transitions, the ratio of column densities and excitation temperatures
$N$/$T_{\rm ex}$ as a function of energy may be described by a
Boltzmann law with a specific rotational temperature $T_{\rm
  rot, JJ'}$ (or simply $T_{\rm JJ'}$)  via

\begin{equation}
\frac{N(J',J')/T'_{\rm {ex}}}{N(J,J)/T_{\rm {ex}}}=\frac{g_{\rm op}(J')}{g_{\rm op}(J)}
\frac{2J'+1}{2J+1}\,\exp\left(\frac{-\Delta E}{T_{\rm JJ'}}\right)
\label{eq:trot}
\end{equation} 
for a pair of inversion lines \amm(J',J') and \amm(J,J) [$g_{\rm
  op}(J)$ is the statistical weight for a given ammonia species, $g_{\rm op}=1$ for
para-ammonia, $g_{\rm op}=2$ for ortho--ammonia].

The assumption of equal $T_{\rm ex}\approx T'_{\rm ex}$ values is justified for
the para--ammonia (1,1), (2,2), and (4,4) lines according to the
radiative transfer models of \citet{wal83} and \citet{dan88}. For the ortho--ammonia
(3,3) and (6,6) inversion transitions, excitation temperatures may
vary within a factor of $\sim 2$ as compared to the lower energy
para--\amm\ transitions.

\begin{figure}
\epsscale{1}
\plotone{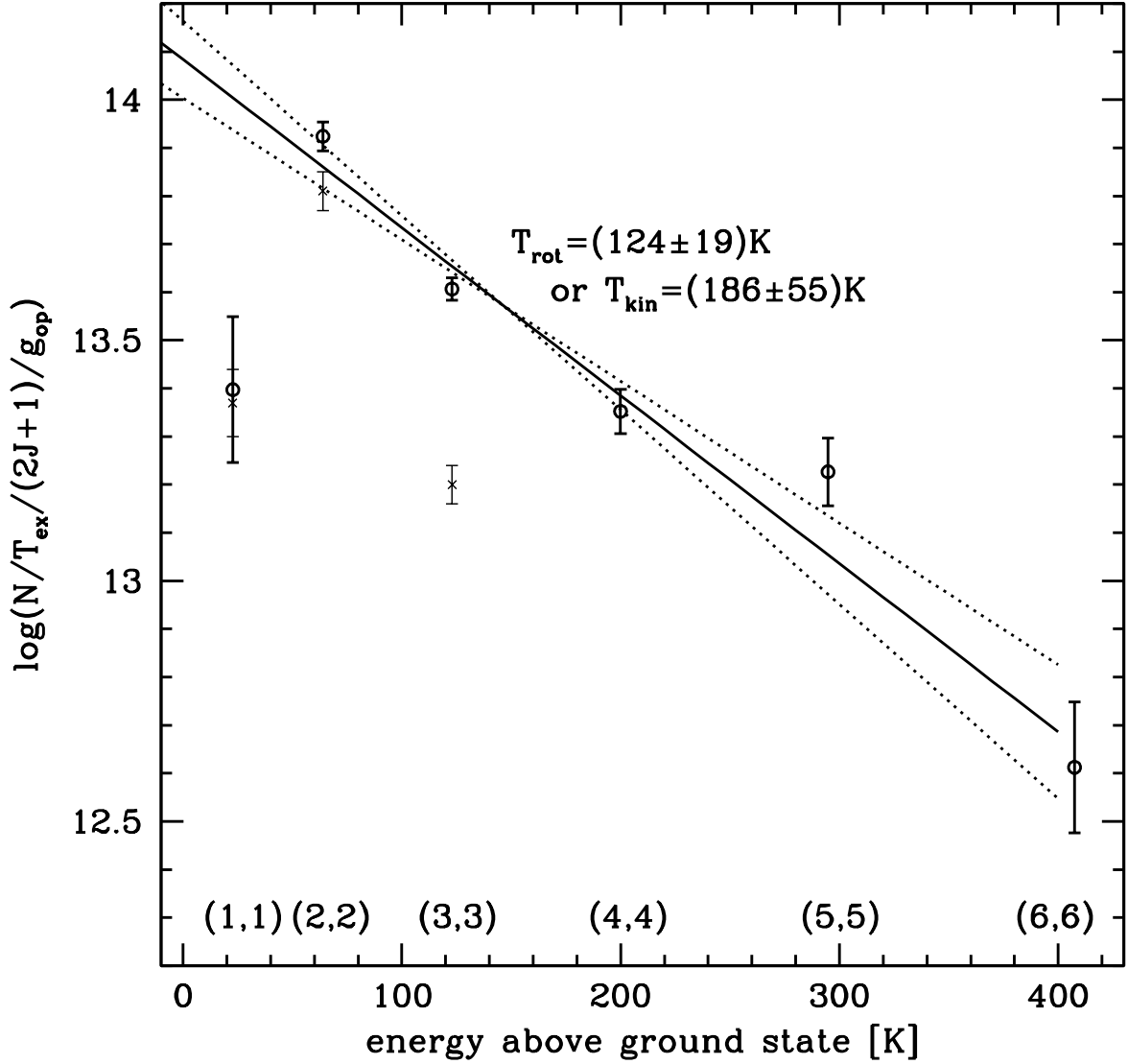}
\caption{Boltzmann plot of the ammonia measurements. The
  ordinate exhibits the weighted logarithm of $N/T_{\rm
    ex}$. We plot the rotational temperature fitted to all but the
  (1,1) datum with the errors in slope. ATCA data points are marked by bold
circles, whereas the 
GBT data is plotted with thin crosses and errorbars. 
  \label{fig:boltz}}
\end{figure}

In Fig.\,\ref{fig:boltz}, we show a Boltzmann plot of the ammonia
inversion lines toward Arp\,220. The GBT and the ATCA data align
reasonably well for the \amm(1,1) and (2,2) transitions. The \amm(3,3)
datum, however, is significantly lower for the GBT observations. The
weighted population of the NH$_3$ (2,2) state is larger than that of
the (1,1) state which is not expected but has been confirmed by our two
independent measurements. The small \amm (1,1) column density is even
more surprising as both transitions, \amm (1,1) and (2,2), belong to
the same para--ammonia species. Both transitions have likely similar
excitation temperatures and are optically thin (Table\,\ref{tab:line},
see also Sect.\,\ref{sec:tau}). This excludes the possibility that
saturation effects are responsible for different strengths of the
(1,1) and (2,2) transitions.

An explanation for the weak \amm(1,1) feature could be that an
additional, cool ammonia component in Arp\,220 contaminates the
absorption spectrum by ammonia (1,1) emission. This cloud needs to be
cool to avoid significant population of the higher energy levels; \amm
(2,2) through (6,6) appear to have absorption profiles that are well
described by a single rotational temperature and are likely little or
not affected by any attenuation from cold gas (see the discussion in
the last paragraph of this Section). To be seen in emission, the cold
component would have to be in the beam but away from the line of sight
toward the continuum. The faint emission channels adjacent to the
(1,1) absorption profile of the ATCA data may provide an indication
for such a cold component.

In spite of this complication and ignoring the unusual profile of the
(1,1) line with respect to all other lines, we can estimate the
rotational temperature of the gas. Using the Boltzmann diagram
(Fig.\,\ref{fig:boltz}), this is performed by a weighted fit through
all (2,2), (3,3), (4,4), (5,5), and (6,6) data points ($T_{\rm
  rot}=T_{23456}$, with an ortho--to--para ammonia abundance ratio of
1). Using the ATCA data, we derive $T_{23456}=124\pm19$\,K and the
best fit is shown in the Boltzmann plot in Fig.\,\ref{fig:boltz}. Note
that the fit describes all data well except for the (1,1) ammonia
transition. The good fit to the (2,2) though (6,6) data provides
confidence that our interpretation and Gaussian fits to the related
spectra are reasonable in terms of peak apparent optical depths and
linewidths. However, it also points toward a less reliable \amm(3,3)
GBT flux measurement. The three lowest transitions of the GBT spectrum
do not line up at all in the Boltzmann plot. In addition, the GBT
\amm(3,3) datum strongly deviates from the best fit of the ATCA
\amm(2,2) through (6,6) data points. We therefore use the superior
number and quality of the ATCA spectra throughout the rest of our
analysis.

Since the lowest energetic ammonia (0,0) transition belongs to the
ortho-ammonia species, cold ammonia {\it formation} temperatures
would be reflected in an enhanced production of ortho-\amm\ over
para-\amm. In that case one would observe the (3,3) and (6,6)
transition data to be significantly elevated from the rotational
temperature fit through all points. A $T_{\rm ex}$ difference between
the ortho and para ammonia inversion lines, however, could compensate
or strengthen a deviation from a single--line fit in the Boltzmann
diagram. Such a difference can be well fitted by a linear function
through \amm(2,2) and higher, which suggests that the
ortho-to-para-\amm\ ratio does not deviate substantially from
unity. In turn, this implies that the ammonia was formed under warm
conditions. \cite{tak02} show that ortho-to-para-\amm\ ratios near
unity require formation temperatures exceeding $\sim 30$\,K.

\begin{figure} 
\epsscale{1.2} 
\includegraphics[angle=0,scale=0.8]{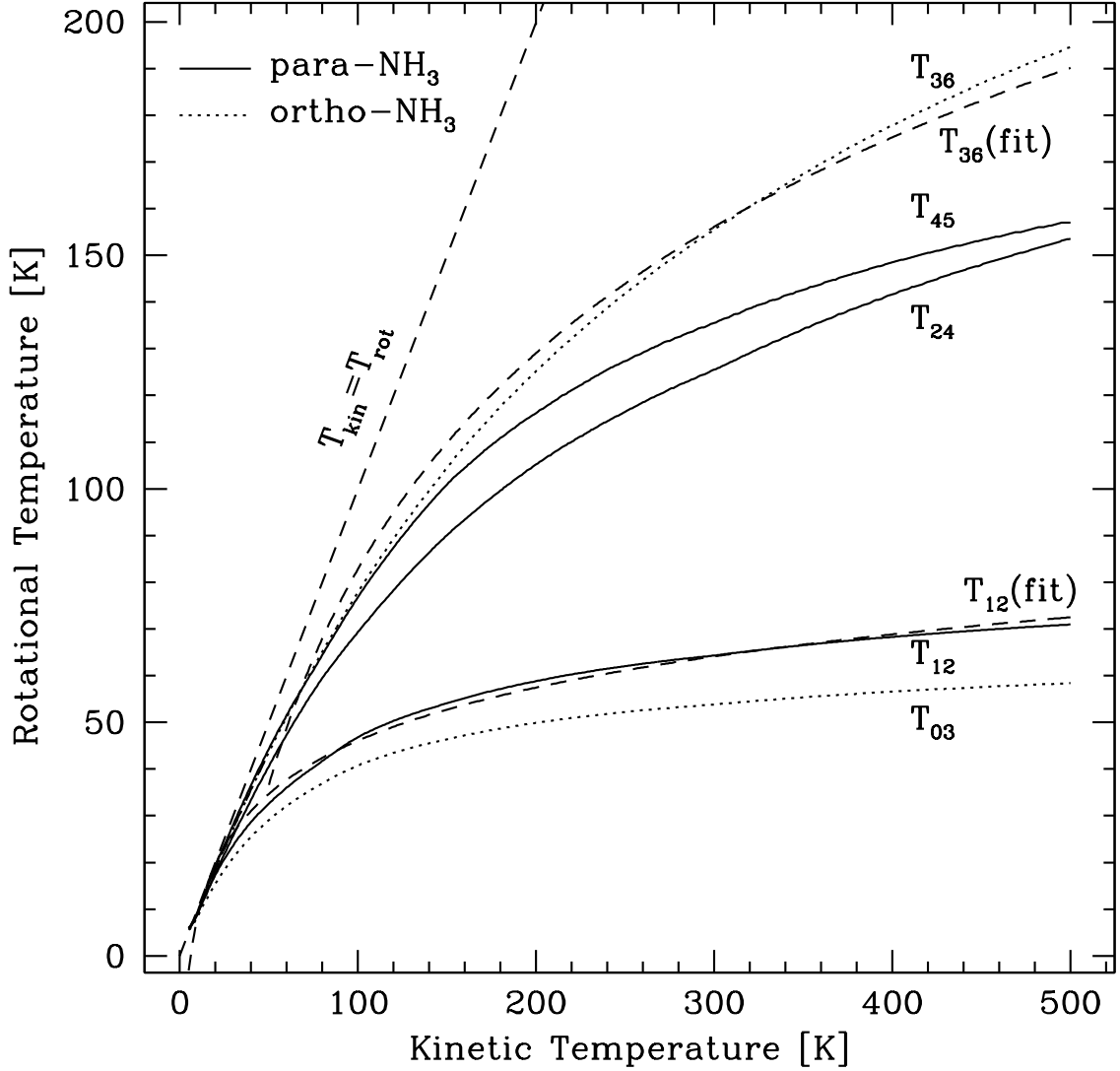} 
\caption{LVG results on the conversion between rotational and kinetic
  ammonia temperatures (solid curves: para-ammonia, dotted curves:
  ortho-ammonia). The subscripts denote the transitions used for the
  $T_{\rm rot}$ determination. The dashed straight line indicates
  where both temperatures would be identical; $T_{\rm rot}$ is always
  underestimating $T_{\rm kin}$. For $T_{\rm 12}$ and $T_{\rm 36}$ we
  show logarithmic fits to the curves in dashed lines. These fits are
  described in eqs.\,\ref{eq:tkintrot12}, and \ref{eq:tkintrot36},
  respectively. \label{fig:tkintrot}}
\end{figure}

For low temperatures, rotational temperatures correspond well to
kinetic temperatures. At higher temperatures, however, populations of
more and more non-metastable levels as well as radiative decay result
in a deviation from the approximation \citet[cf.][]{wal83}. In
Fig.\,\ref{fig:tkintrot} we show the results on the $T_{\rm
kin}$--$T_{\rm rot}$ relation for different pairs of inversion
transitions. This calculation is based on an LVG radiative transfer
model (described in \citealt{ott05}, using the collision rates of
\citealt{dan88}). For the \amm(1,1) and (2,2) transitions, $T_{\rm
kin}\approx T_{\rm 12}$ is valid below $\sim 40$\,K.Using higher
transitions, the approximate equality is valid into progressively
higher temperature regimes. For that reason, it is preferable to
observe high \amm\ transitions if they are bright
enough. Nevertheless, even if the $T_{\rm kin}\approx T_{\rm rot}$
relation is broken, one can still apply the models displayed in
Fig.\,\ref{fig:tkintrot} as they only weakly depend on volume density
\citep[for details, see][]{ott05}. Here we present logarithmic fits of
$T_{\rm kin}$ to the rotational temperatures $T_{\rm 12}$ and $T_{\rm
36}$, that are based on the lowest para-ammonia (1,1) and (2,2), and
the lowest ortho-ammonia (3,3) and (6,6) transitions, respectively:

\begin{equation}
T_{\rm kin} = 6.06\times \exp(0.061 \; T_{\rm 12}).   
\label{eq:tkintrot12}
\end{equation}

and 

\begin{equation}
T_{\rm kin} =
\begin{cases}
T_{\rm  36} & \text{for $T_{\rm  36}\lesssim 50$\,K}\\
28.9 \; \times \exp(0.015 \; T_{\rm 36})   & \text{for $T_{\rm  36}\gtrsim 50$\,K.}
\end{cases}
\label{eq:tkintrot36}
\end{equation}
\placefigure{fig:tkintrot}

\noindent Both fits to the LVG solutions are shown in
Fig.\,\ref{fig:tkintrot} as dashed lines to their corresponding
curves.

For Arp\,220, we determine the rotational temperature across five
transitions, rather than only a pair. Eq.\,\ref{eq:tkintrot12} cannot
be applied as the (1,1) level seems to be an outlier in Arp\,220 and
was excluded from our $T_{\rm rot, 23456}$ fit.  We are using
Eq.\,\ref{eq:tkintrot36} instead. This rotational temperature is
calculated over a much larger energy difference of the (3,3) and (6,6)
levels, rather than the smaller difference between (1,1) and (2,2). In
addition, a Boltzmann fit through (3,3) and (6,6) provides the same
124\,K rotational temperature as the fit through (2,2)-(6,6). For that
reason, we expect that the conversion of Eq. \,\ref{eq:tkintrot36} is
a good approximation for the kinetic temperature of the ammonia-traced
gas in Arp\,220 and we derive $T_{\rm kin}=(186\pm55)$\,K. 

Such a kinetic temperature is similar to that found in the more nearby
`normal' starburst galaxy NGC\,253, which exhibits $T_{\rm
kin}=130-210$\,K \citep{ott05}. Other starburst galaxies, such as
Maffei\,2, and M\,82 \citep[][]{hen00,wei01,tak02,mau03} show lower
rotational temperatures than NGC\,253, which must also result in lower
kinetic temperatures.

\begin{figure}
\epsscale{1}
\plotone{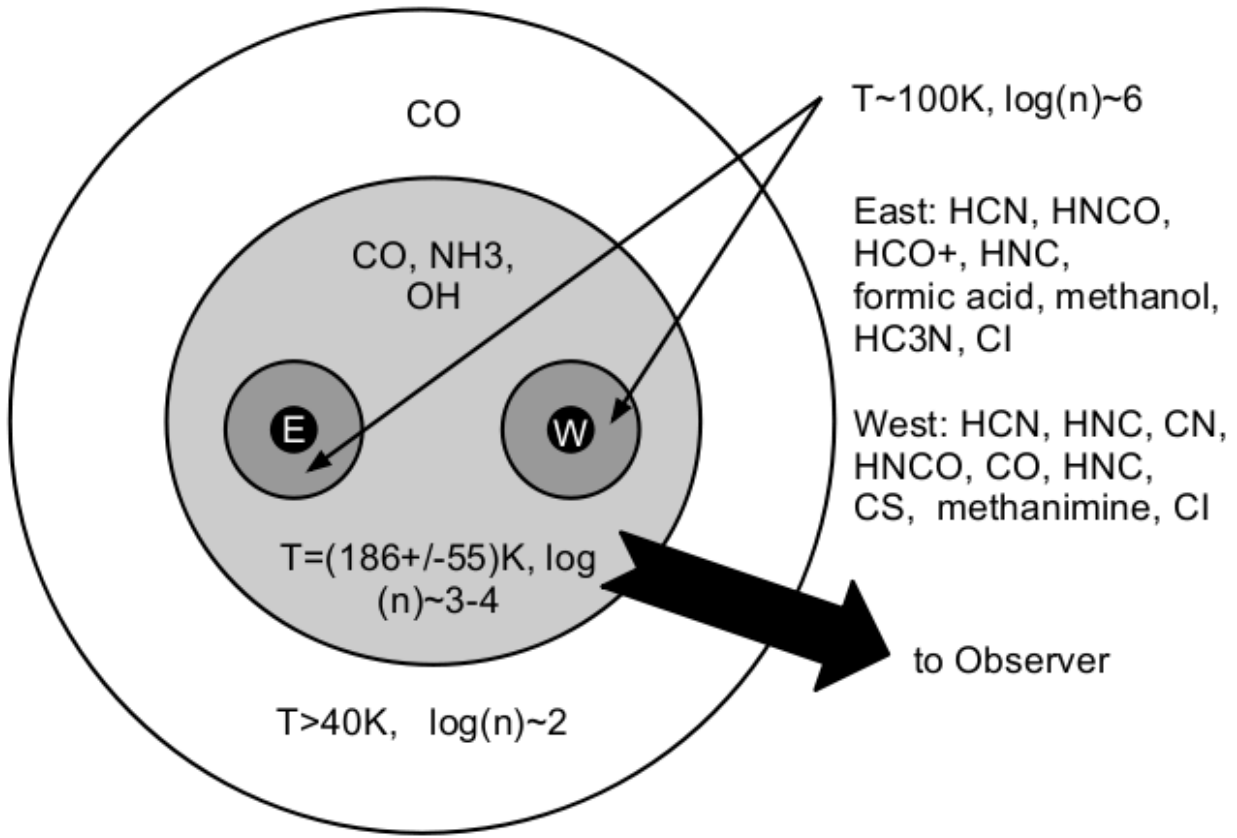}
\caption{A cartoon image of Arp\,220. Shown are the different gas
  phases and some characteristic molecular or atomic tracers. The
  outer envelope has a lower limit to the temperature and the \amm\
  component likely resides between the outer envelope and the inner
  dense regions that surround the eastern and western
  cores. \label{fig:cartoon}}
\end{figure}

For comparison, \citet{dow07} report an intrinsic hot dust temperature
component with $T_{\rm dust}=170$\,K toward the western nucleus of
Arp\,220. This is comparable to what we obtain for \amm, but might not
emerge from the same physical feature (see Sect.\,\ref{sec:kin}). Dust
temperatures are somewhat lower in the outer molecular ring and over
the entire body of Arp\,220. {\it ISO} observations show two dust
temperature components with 47 and 120\,K \citep[][]{kla97}. The gas
thus has a kinetic temperature component as high as that of the
hottest dust component in Arp\,220. Recent SMA
\citep{sak08,aal09}, ASTE \citep{ima10}, and JCMT data
\citep{gre09,pap10} of highly excited, dense gas show typical temperatures
of $\sim 45-120$\,K in the densest, $\sim 10^{6}$\,\vden\ components,
likely the gas closest to the cores. The temperature of the gas phase
we determine is somewhat higher and may arise within the surrounding
ring or disk (see Sects.\,\ref{sec:dens} and \ref{sec:kin}, see also
Fig.\,\ref{fig:cartoon}).  No direct temperature measurement has so
far been reported for this physical feature.

Given the rotational temperatures, we can now estimate properties of
any ammonia that contributes emission to explain the low absorbed flux
of the (1,1) line relative to the (2,2) line. Following the Boltzmann
plot (Fig.\,\ref{fig:boltz}), and assuming that the derived
$T_{23456}$ rotational temperature can be extrapolated to a
hypothetical (1,1) column density, one would expect a $\sim 4$ times
larger column density of the (1,1) line than actually measured in
absorption. According to Eq.\,\ref{eq:NT}, this would result in an
integrated apparent optical depth $\sim 4$ times larger. This missing
integrated apparent optical depth must be compensated for by emission
of order $\sim 20$\,mJy. The gas responsible for such emission would
have to be cold to avoid contaminating the other inversion
profiles. For an upper limit, let us assume that the \amm(2,2) line is
uncontaminated at a level two times below the one actually measured
and that, from that level, the additional emission component raises
the (negative) flux to the measured value. The emission line would
then require a strength of $\sim 6$\,mJy. Using this value as an upper
limit to the \amm(2,2) emission component, we can derive an upper
limit to the temperature from the (1,1) and (2,2) lines. Using the
equations for ammonia emission as provided in \citet{ott05}, we
estimate an upper rotational temperature limit of $T_{\rm
12}\lesssim20$\,K. With Eq.\,\ref{eq:tkintrot12}, this translates to
an upper limit of the kinetic temperature of the same order of $T_{\rm
kin}\lesssim20$\,K. This scenario can be proven by observations at
high spatial resolution ($\lesssim 0.5\arcsec$) that allow direct
imaging of emission components offset to the lines of sight toward the
bright continuum sources. The EVLA may provide such observations in
its extended configurations. Alternatively, mm/submm interferometers
such as ALMA and PdBI may be able to confirm the existence of this
cold component. The ammonia inversion lines, however, do not fall in
their wavebands and the predicted cold gas component would need to be
revealed via low energy, high frequency transitions of other suitable
molecular tracers.

\subsection{Gas Densities}
\label{sec:dens}

The excitation temperatures of ammonia in ULIRGs is still unclear. An
indication of a reasonable range might be given by the calculations of
\citet{wal83} (see their fig.\,3, and using their own collision
rates). For the ratio of an \amm\ concentration to velocity gradient
of 10$^{-4}$\,cm$^{-3}$\,(\kms\,pc$^{-1}$)$^{-1}$, they show that the
excitation temperatures of the ammonia inversion lines span $T_{\rm
ex}\sim 10-50$\,K, except for a poorly confined range of parameters
where the ammonia (3,3) line exhibits maser emission. All of the
inversion states have similar excitation temperatures at a given
density. ULIRGs such as Arp\,220, however, could show substantially
different excitation temperatures due to their high gas densities,
high kinetic temperatures, and strong velocity
gradients. Nevertheless, assuming that the \citet{wal83} calculations
provide at least an indication of the possible range, we can derive
ammonia column densities using Eq.\,\ref{eq:NT}. The results are
listed in Table\,\ref{tab:dens}. The individual inversion states
exhibit column densities of order $\sim10^{16}$\,\cden\ for $T_{\rm
ex}=50$\,K and $\sim10^{15}$\,\cden\ for $T_{\rm ex}=10$\,K, which add
up to $(7.8\pm0.4)\times10^{16}$\,\cden\ and
$(1.6\pm0.1)\times10^{16}$\,\cden\ for the two excitation temperature
values, respectively. The relation scales linearly with $T_{\rm ex}$,
so even a high value of 100\,K or 150\,K would only modestly increase
the column densities.  The (0,0) ortho-ammonia level is degenerate and
does not exhibit an inversion line. However, it is the lowest energy
state of ammonia and thus has a significant population. To estimate
the amount of ammonia in the (0,0) state, we extrapolate the $T_{\rm
23456}=(124\pm19)$\,K rotational temperature in the Boltzmann diagram
(Fig.\,\ref{fig:boltz}; assuming an ortho-to-para-\amm\ ratio of
unity) and derive (0,0) columns of $(6.0\pm0.9)\times10^{15}$\,\cden\
and $(1.2\pm0.2)\times10^{15}$\,\cden\ for excitation temperatures of
50\,K and 10\,K, respectively. Adding this to the (1,1) through (6,6)
columns, we derive total ammonia columns of
$(8.4\pm0.5)\times10^{16}$\,\cden\ and
$(1.7\pm0.1)\times10^{16}$\,\cden\ for values of $T_{\rm ex}=50$\,K
and 10\,K, respectively (see Table\,\ref{tab:dens}). This value
neglects all higher level populations, a valid assumption given the
negligible contributions that these will make to the entire column if
the rotational temperature stays the same for $J>6$. For example, the
(7,7) inversion line is 538\,K above the ground state; by
extrapolating $T_{\rm rot}$ via Eq.\,\ref{eq:trot} in the Boltzmann
diagram (Fig.\,\ref{fig:boltz}), we see that the \amm(7,7) line would
fall at a $N/T_{\rm ex}$ that is $\sim 14$ times below the
\amm(6,6) value.

As shown in Fig.\,\ref{fig:gbtspec} and in Table\,\ref{tab:line}, some
peak and integrated apparent optical depths differ between the ATCA
and GBT observations. In the following, we try to derive how the
properties of the gas have changed if the different spectra are due to
a time variation of the gas in front of the line of sight to the
continuum (the GBT data were observed about a year after the ATCA
data). To start, we use Eq.\,\ref{eq:NT} to determine $N/T_{\rm ex}$,
which is directly proportional to the integrated apparent optical
depths. According to Table\,\ref{tab:line}, this translates to ATCA
over GBT column density ratios of $\sim1.3$ and $2.6$ for the \amm
(2,2) and (3,3) transitions, respectively, at a given $T_{\rm
ex}$. The total ammonia columns depend on the temperature assumed,
but, as shown in Fig.\,\ref{fig:boltz}, the GBT data points are not
reliable enough to determine a temperature as the \amm(1,1) line has
very different properties. If the (2,2) and (3,3) lines are
representative of the gas phase measured by the GBT, however, we can
estimate a rotational temperature from the slope and we derive a cold
$T_{\rm 23}\sim 42$\,K. Extrapolating this to higher transitions via
Eq.\,\ref{eq:trot} and adding the \amm(1,1) will then provide $\sim
28\times10^{15}$\,\cden\ for $T_{\rm ex}=50$ and $\sim
6\times10^{15}$\,\cden\ for $T_{\rm ex}=10$, or $\sim 2.5$ times less
than what we estimate from the ATCA data (see
Table\,\ref{tab:dens}). This is the variation of the ammonia column
that potentially may have modified the absorption profile between the
ATCA and GBT measurements over the time of a year.

Eventually, we can use the total column densities to derive ammonia
volume densities. \citet{dow98} show that the bulk of the molecular
gas in Arp\,220 is found in a ring or disk (the light gray shaded
area in Fig.\,\ref{fig:cartoon}). In their model, the size of this
ring or disk is about $\sim 700$\,pc when measured from the western
edge closest to the observer toward the eastern nucleus, the farthest
continuum source against which we see the \amm\ absorption. This
geometry leads to \amm\ volume densities of $f_{\rm
  V}^{-0.5}\times(3.9\pm0.2)\times10^{-5}$\,\vden\ and $f_{\rm
  V}^{-0.5}\times(0.8\pm0.1)\times10^{-5}$\,\vden\ for the $T_{\rm
  ex}=50$\,K and 10\,K limits, respectively ($f_{\rm V}$ is the volume
filling factor of the dense gas). If the line of sight through the
molecular gas is substantially smaller, i.e. if the absorption occurs
mainly against the western core of Arp\,220 (see
Sect.\,\ref{sec:kin}), the volume densities increase by a factor of
$\sim 3$.

We can now estimate the volume densities of molecular hydrogen that is
traced by the ammonia absorption lines. To do so, we have to assume a
fractional abundance of ammonia. \citet{ott05} and \citet{mau03} show
that for a sample of nearby, non-ULIRG starburst galaxies, $[N({\rm
  NH_{3}})/N({\rm H_{2}})]$ hovers around $10^{-8}$. If this value is
applicable to the dense gas in Arp\,220, the \mh\ densities are of
order $f_{\rm V}^{-0.5}\times (1-4)\times10^{3}$\,\vden. For a filling
factor of unity, this is similar to the environment of the Sgr\,B2
region near the Galactic Center \citep{hue95b} and close to the
critical density of ammonia which is $\sim 2\times10^{3}$\,\vden\
\citep{roh04}. \citet{sco97}, however, estimate a density of $\sim
2\times10^{4}$\,\vden\ for the molecular ring in Arp\,220. If the
ammonia absorption lines trace the same gas, the volume filling factor
would be of order $f_{\rm V}\sim 0.01$.

In Sect.\,\ref{sec:T}, we speculate that the small absorption depth of
the (1,1) line relative to the other \amm\ lines might be caused by an
additional, cold layer of ammonia in Arp\,220. The component would
have to be located inside the beam but should be displaced from the
lines of sight toward the continuum sources and thus be seen in
emission. If this line has a width of $\sim 200$\,\kms, such as to
compensate some of the absorption line, the assumed strength can be
converted to a column density in the (1,1) line of $\sim
1\times10^{14}$\,\cden.  With $T_{\rm 12}=20$\,K (Sect.\,\ref{sec:T})
and using the Equation (A15) in \citet{ung86}, this can be
extrapolated to a total column of $N_{\rm
  emission}\sim2\times10^{14}$\,\cden. The hypothetical emission
component thus has a column density more than 2 orders of magnitude
smaller than the ammonia layer that is seen in absorption against the
background continuum.

\subsection{Gas Kinematics}
\label{sec:kin}

The ammonia linewidths, $\sim 230-430$\,\kms\ (\amm(1,1): 121\,\kms),
are at the upper end and even wider than those found in nearby
`normal' starburst galaxies, such as NGC\,253 \citep[$\sim
250$\,km\,s$^{-1}$;][]{ott05}, Maffei\,2 ($\sim 150$\,km\,s$^{-1}$),
or IC\,342 \citep[$\sim 50$\,km\,s$^{-1}$;][]{mau03} when averaged
over the nuclear regions. Toward Arp\,220, molecular emission lines
such as CO and HCO$^{+}$ and HNC \citep[e.g., see the summary in
][]{gre09} tend to be broadened to $\sim 320-430$\,km\,s$^{-1}$, as
are highly excited CO and CS transitions
\citep[e.g.]{sak08,gre09,pap10} which show linewidths of $\sim
300$\,\kms\ for each of the nuclear components. Imaging of the CO
($J=2\rightarrow1$) distribution with sub--arcsecond resolution
\citep{sco97,dow98} provides information on the dynamical properties
of the molecular gas. Toward the east of the active region that
encompasses the two galactic nuclei, the velocity range is
$5500-5800$\,\kms\ and toward the west it is $5200-5500$\,\kms\
\citep{sco97}. One possibility is that the molecular gas we observe in
ammonia, $\sim 5450$\,\kms\ with a linewidth of $\sim 230-430$\,\kms,
has components toward both nuclei. This, however, is unlikely since
the absorption spectra are well defined by a one-component Gaussian
that cannot be explained by two disks, separated in velocity space
well above our velocity resolution. Alternatively, the width may
reflect the difference in systemic velocities of the cores, which is
$\sim 300$\,\kms\ \citep[e.g.][]{dow98}. It is a coincidence that the
linewidths of the gas around the individual cores and the velocity
difference between the cores is of the same order. This makes it
difficult to determine the origin of the gas based on kinematic
information alone. The \amm\ spectra show similarities to previously
measured OH lines. The absorption lines of rotationally excited OH
exhibit linewidths of $\sim 250-370$\,km\,s$^{-1}$
\citep{hen86,hen87,sal08} which is like the linewidths obtained from
\amm\ and, as \amm, OH absorption has no clear velocity imprint of
either nucleus. Both ammonia and OH may therefore trace the same
molecular material either from a single nucleus, such as the linewidth
suggests, or, alternatively, in the common disk in which the two
nuclei are embedded, as suggested by the systemic velocities.

The gas in the dense cores has volume densities of $\sim
10^{6}$\,\vden\ (see Sect.\,\ref{sec:dens}), but we estimate a 2-3
orders of magnitude lower density for the \amm\ traced gas
(Table\,\ref{tab:dens}). A combination of the systemic line
velocities, the derived densities, and the temperature argues that the
ammonia- (and OH-) traced gas is more likely to be stored in the inner
ring or disk that surrounds both nuclei (see Fig.\,\ref{fig:cartoon}).
Unfortunately, as for ammonia, only single-dish data of rotationally
excited OH have been published to date and its distribution with
respect to the nuclei in Arp\,220 has not yet been directly
imaged. Sub-arcsecond mapping of the components with the EVLA or ALMA
is needed to at least determine the absorption profiles against each
nucleus. Such observations will be crucial to obtain a better model of
the layered gas distribution in Arp\,220.

\section{Summary}
\label{sec:summary}

In this article, we have presented ATCA and GBT observations of ammonia
toward the ultraluminous far-infrared galaxy Arp\,220. We conclude the
following:

\begin{enumerate}

\item We detect all observed ammonia lines ($J$,$K$)=(1,1), (2,2),
  (3,3), (4,4), (5,5), and (6,6), toward the bright continuum in
  absorption.

\item The detected ammonia linewidths are in the range $\sim
  120-430$\,\kms, in agreement with those of other molecules. We
  cannot confirm the extremely wide lines reported by
  \citet{tak05}. 

\item Based on the ATCA ammonia data (but excluding
  the \amm(1,1) line), we derive the rotational temperature of
  the molecular gas to be $(124\pm19)$\,K. This translates to a
  kinetic temperature of $T_{\rm kin}=(186\pm55)$\,K. The kinetic
  temperature of \amm\ is similar to that found in the nucleated
  starburst galaxy NGC\,253.

\item We see no evidence for an ortho--to--para-ammonia ratio other
  than unity. This implies ammonia formation temperatures exceeding
  $\sim 30$\,K.

\item Combined column densities of the measured lines plus the
  extrapolated (0,0) column add up to
  $(1.7\pm0.1)\times10^{16}$\,\cden\ and
  $(8.4\pm0.5)\times10^{16}$\,\cden\ for assumed excitation
  temperatures of 10\,K and 50\,K, respectively. We estimate the
  volume density of the molecular gas that is traced by ammonia to be
  $\sim f_{\rm V}^{-0.5}\times(1-4)\times 10^{3}$\,\vden\ ($f_{\rm
    V}$: volume filling factor). Together with the temperature
  estimate and the fact that the spectra are well described by
  single-component Gaussians, we conclude that the bulk of this
  moderately dense gas likely surrounds both starburst nuclei.

\item Widespread gas, based on low excitation CO, exhibits densities
  that are about two orders of magnitude less than the ammonia-traced
  gas layer. This component defines the outer disk of Arp\,220. In
  contrast, the gas close to the two starburst nuclei is characterized
  by densities two orders of magnitude larger than the ammonia-traced
  gas. The gas that we see via ammonia is thus likely a layer that is
  located between the nuclei and the outer envelope. The geometry is
  most likely a ring or disk that surrounds both nuclei. The
  temperature profile appears to have a local maximum at the position
  of the gas that is traced by ammonia.

\item The \amm(1,1) line is weaker than expected for thermal gas based
  on all other measured \amm\ transitions. A possible explanation for this
  unusual behavior could be an additional, cold, $\lesssim 20$\,K
  component of molecular gas, located inside the
  beam but displaced from the lines of sight toward the
  continuum sources in Arp\,220. This gas, if present, would have to
  be at an estimated column density of $N_{\rm
    emission}\sim2\times10^{14}$\,\cden\ to sufficiently mask the
  (1,1) absorption. The cool temperature is required to keep the
  masking of the other lines to a minimum.

\item We also detect an absorption feature adjacent in frequency to
  the \amm(3,3) line in the ATCA data but not in the GBT data. This
  absorption feature may be interpreted as an OH $^{2}\Pi_{3/2}$
  $J=9/2$ $F=4-4$ transition, or the combination of the F=4-4 and
  F=5-5 OH doublet. Both options yield line parameters similar to that
  of ammonia, including a rotational temperature that is about 2 times
  larger for OH. If this detection holds, it would be the first
  extragalactic evidence for this line which has so far only been seen
  toward Galactic compact \ion{H}{2} regions such as W3(OH).

\item The ATCA and GBT data show some differences in their line
  shapes. This is most likely an instrumental effect of the two very
  different telescopes and observing procedures. We cannot exclude,
  however, that there is intrinsic time variability of the gas
  distribution in Arp\,220 along the narrow lines of sight toward the
  absorbing background structure. If this is the case, then the GBT
  data suggests gas with a $\sim 3$ times cooler rotational
  temperature than that of the ATCA measurement, and the column
  density would be $\sim 2.5$ times lower at the time when the GBT
  observations were taken (a year after the ATCA observations). Given
  the typical cloud and continuum size scales of a few to tens of
  parsec and the velocities of a few hundred \kms, however,
  drastic variations of the absorption spectra on annual timescales
  seem to be unlikely and an instrumental effect is more plausible. 

\end{enumerate}

\acknowledgments
We thank Fabian Walter for his comments on the manuscript.
C. H. thanks ATNF for support during his time spent at ATNF
and ATCA. This research has made use of the NASA/IPAC
Extragalactic Database (NED), which is maintained by the Jet
Propulsion Laboratory, Caltech, under contract with the National
Aeronautics and Space Administration (NASA) and NASA's Astrophysical
Data System Abstract Service (ADS).

Facilities: \facility{ATCA,GBT}.


\clearpage

\clearpage

\clearpage

\clearpage

\clearpage

\clearpage

\clearpage
\begin{deluxetable}{lllllll}
\tabletypesize{\scriptsize}
  \tablecaption{ Observational Parameters. All observations were
    directed toward RA(J2000)=15$^{h}$34$^{m}$58\fs10,
    DEC(J2000)=23\degr30\arcmin11\farcs0. For the assumed 75\,Mpc
    distance of Arp\,220, 1\arcsec\ corresponds to 0.36\,kpc. \label{tab:obs}} 
  \tablewidth{0pt}
  \tablehead{ 
     \colhead{Telescope} &  \colhead{Date} & \colhead{Center
       Frequencies} & \colhead{BW\tablenotemark{a}} &
     \colhead{$\Delta \nu$\tablenotemark{b}} & \colhead{Beam size} & \colhead{\amm\
       Transition}\\
     \colhead{} &  \colhead{} & \colhead{[GHz]} & \colhead{[MHz]} &
     \colhead{[MHz]} & \colhead{[\arcsec]} & \colhead{}}
\startdata
\hline
ATCA & 2005 Jul 28-31, Aug 01 & 23.14, 23.20, 23.26, 23.32
& 128& 4.4 & $31.1\times21.1$& (1,1), (2,2), (3,3) \\
&& 23.38, 23.44, 23.50, 23.56 & &&& \\
ATCA & 2005 Jul 30 & 23.665, 23.737 &128 & 4.4 & $31.2\times21.4$& (4,4)\\ 
ATCA & 2005 Jul 31 & 24.052, 24.124 &128 & 4.4 &
$30.9\times19.8$& (5,5)\\ 
ATCA & 2005 Mar 17 and 20 & 24.603 &128 & 4.4 & $10.7\times7.7$& (6,6)\\ 
GBT & 2006 Jan 21 and Feb 02 & 23.445 & 800 & 1.56 & 33 &  (1,1), (2,2), (3,3) \\
\hline
        
\enddata

\tablenotetext{a}{BW: bandwidth of each frequency setting}
\tablenotetext{b}{$\Delta\nu$: spectral resolution}

\end{deluxetable} 

\clearpage

\begin{deluxetable}{lcccc}
  \tablecaption{Ammonia (1,1) through (6,6) line
    parameters. Velocities are given with respect to the redshift of
    Arp\,220 of $z=0.018126$, or a velocity of 5434\,\kms. The
    parameters of the OH $^{2}\Pi_{3/2}$ $J=9/2$ $F=4-4$ line
    candidate are also provided. \label{tab:line}} \tablewidth{0pt}
  \tablehead{ \colhead{\amm\ line} & \colhead{$\tau^{\rm peak}$} &
    \colhead{$v_{\rm opt,bary}^{\rm peak}$} & \colhead{$\Delta v_{1/2}
      $} &
    \colhead{$\int \tau\, $d$v$. }\\
    \colhead{} & \colhead{} & \colhead{[\kms]} & \colhead{[\kms]} &
    \colhead{[\kms]} } \startdata

\multicolumn{5}{c}{ATCA}\\

\hline

(1,1) & $    0.046 \pm0.012  $ & \phs$63  \pm   16  $ & $    121
\pm  38  $ & $    5.9 \pm  2.5  $ \\
           
(2,2) & $   0.178 \pm 0.008 $ & $-39  \pm    5  $ & $     232 \pm
13 $ & $   43.9 \pm  3.2$ \\
           
(3,3) & $   0.145 \pm 0.005 $ & \phn$-1\pm     8  $ & $    435 \pm    19 $ & $  67.0  \pm 3.7 $ \\           

(4,4) & $ 0.085 \pm 0.006 $ & $ -18 \pm 10 $ & $ 285 \pm 25 $ & $ 25.9
\pm 2.9 $ \\ 

(5,5) & $0.072 \pm 0.008 $ & \phs$ 45 \pm 18 $ & $ 327 \pm 45 $ & $ 25.1
\pm 4.4 $\\
 
(6,6) & $ 0.53 \pm 0.016 $ & \phs$ 20 \pm 55 $ & $ 268 \pm 56 $ & $15.1\pm5.6$\\

\hline

OH F=4-4 & $    0.086 \pm 0.006 $ & $     -13   \pm     10  $ & $
292  \pm  29 $ & $   26.7 \pm  3.2 $ \\

\hline
\multicolumn{5}{c}{~}\\
\multicolumn{5}{c}{GBT}\\

\hline

(1,1) & $    0.042 \pm 0.005  $ & \phs$     27  \pm   7  $ & $    121
\pm  17  $ & $  5.4  \pm 1.0  $ \\
           
(2,2) & $  0.136 \pm 0.004 $ & $     -35  \pm   10  $ & $     230 \pm 22
$ & $ 33.3 \pm 3.3 $  \\
           
(3,3) & $   0.094 \pm 0.003 $ & $      -42  \pm   10  $ & $    258 \pm
22 $ & $  25.8  \pm 2.4 $ \\    
        
\enddata

\end{deluxetable}

\clearpage

\begin{deluxetable}{lcc}
  \tablecaption{ Densities of the molecular gas. The center column is
    derived using an excitation temperature of 50\,K and the right
    column using $T_{\rm ex}=10$\,K. Whereas $\sum_{\rm J=1...6}\,
    N(J,J)$ denotes the sum of the measured lines, the ammonia column
    densities listed as $\sum_{\rm J=0...6}\, N(J,J)$ include the
    (0,0) inversion state that was extrapolated using $T_{\rm
      rot}=124$\,K. The column densities transform into volume
    densities $n_{\rm NH_{3}; J=0...6}$ for an assumed line of sight
    of 700\,pc. Finally, for a fractional abundance of $[N({\rm
      NH_{3}})/N({\rm H_{2}})]=10^{-8}$, the resulting molecular
    hydrogen densities $n_{\rm H_{2}}$ are shown in the last
    row. $f_{\rm V}$ is the volume filling factor.
\label{tab:dens}}
\renewcommand{\arraystretch}{1.2}
\tablewidth{0pt}
\tablehead{
\colhead{} & \colhead{$T_{\rm ex}=50$\,K} & \colhead{$T_{\rm ex}=10$\,K} \\
}
\startdata
&\multicolumn{2}{c}{$N$ [$10^{15}$\,\cden]}\\
\amm (1,1) &  $3.7\pm1.6$  & $0.7\pm0.3$ \\
\amm (2,2) &  $21.0\pm1.5$ & $4.2\pm0.3$ \\
\amm (3,3) &  $28.3\pm1.6$ & $5.6\pm0.3$ \\
\amm (4,4) &  $10.1\pm1.1$ & $2.0\pm0.2$ \\
\amm (5,5) &  $9.3\pm1.6$ & $1.9\pm0.3$ \\
\amm (6,6) &  $5.3\pm2.0$  & $1.1\pm0.4$ \\
\amm(0,0)$_{extr}$ & $6.0 \pm 0.9 $ & $ 1.2 \pm 0.2 $ \\
\tableline
$\sum_{\rm J=1...6}\, N(J,J)$ & $78\pm4$ &  $15.5\pm 0.9$\\
$\sum_{\rm J=0...6}\, N(J,J)$ & $84\pm5$ &  $16.7\pm 1.0$\\
\tableline
& \multicolumn{2}{c}{$f_{\rm V}^{-0.5}\times n_{\rm NH_{3}; J=0...6}$  [$10^{-5}$\,\vden]}\\
&  $3.9\pm0.2$ & $0.8\pm0.1$\\
\tableline
& \multicolumn{2}{c}{$f_{\rm V}^{-0.5}\times n_{\rm H_{2}}$ [$10^{3}$\,\vden] }\\
& $3.9\pm0.2$ & $0.8\pm0.1$\\
\enddata

\end{deluxetable}

\end{document}